\documentclass{llncs}
\usepackage[utf8]{inputenc}
\usepackage{microtype}
\usepackage{textcomp}

\usepackage{listings}
\usepackage{xcolor}
\usepackage{booktabs}
\usepackage{multirow}
\usepackage{amsmath,amssymb}
\usepackage{textgreek}

\usepackage{xcolor}
\newcommand\NEW{}
\newcommand\NEEW{}
\newcommand\NEEEW{}
\newcommand\NEEEEW{}
\newcommand\newml[1]{#1}

\usepackage{hyperref}
\hypersetup{colorlinks,urlcolor=blue,citecolor=blue,linkcolor=blue}

\usepackage{pdfpages}

\title{Improving the Performance of the Paisley~Pattern-Matching EDSL by
  Staged~Combinatorial Compilation}
\author{Baltasar Trancón y Widemann \and Markus Lepper\\\texttt{baltasar@trancon.de}}
\institute{semantics GmbH, Berlin, DE}

\begin{document}
\maketitle

\begin{abstract}
  {\sffamily Paisley} is a declarative lightweight embedded domain-specific
  language for expressive, non-deterministic, non-invasive pattern matching on
  arbitrary data structures in Java applications.  As such, it comes as a pure
  Java library of pattern-matching combinators and corresponding programming
  idioms.  While the combinators support a basic form of self-optimization
  based on heuristic metadata, overall performance is limited by the distributed
  and compositional implementation that impedes non-local code optimization.  In
  this paper, we describe a technique for improving the performance of
  {\sffamily Paisley} transparently, without compromising the flexible and
  extensible combinatorial design.  By means of distributed bytecode generation,
  dynamic class loading and just-in-time compilation of patterns, the run-time
  overhead of the combinatorial approach can be reduced significantly, without
  requiring any technology other than a standard Java virtual machine and our
  {\sffamily LLJava} bytecode framework.  We evaluate the impact by comparison
  to earlier benchmarking results on interpreted {\sffamily Paisley}.  The key
  ideas of our compilation technique are fairly general, and apply in
  principle to any kind of combinator language running on any jit-compiling
  host.
\end{abstract}

\section{Introduction}
\label{intro}

In declarative programming languages with algebraic datatypes,
\emph{constructing} and \emph{querying} structured data are symmetric tasks,
handled by languages features of equal expressiveness, the latter namely by
\emph{pattern matching}.  Semantics are given by a clean, reversible
\emph{algebraic} interpretation.
In object-oriented languages, by contrast, the query side is markedly deficient
in expressiveness~\cite{mowp,paisley-atps}.  This is due partly to shortcomings in
language design, partly to the doctrine of \emph{data abstraction} which is
generally incompatible with algebraic semantics.
\textsf{Paisley}\,\cite{paisley-icmt} is a solution for this dialectic problem.
It is a lightweight embedded domain-specific language (EDSL) that raises the
pattern-matching expressiveness of the host language Java considerably, without
breaking either the imperative control flow or the abstraction of
object-oriented data models.

{\NEEEEW
The present paper summarizes the design of \textsf{Paisley} in
section~\ref{intro}.  Its main contribution is the description and
evaluation of a novel compilation technique, presented in sections
\ref{compile} and \ref{eval}, respectively.}

\subsection{Basic Design and Usage of Paisley}

The lightweight implementation of the \textsf{Paisley} EDSL is
 a pure Java library that runs on a vanilla Java platform
requiring neither compiler nor runtime extensions, and that \emph{reifies}
pattern matching primitives by a collection of Java classes.  Constructor terms
for objects of these classes form a declarative language,
but since they {\NEEEEW denote} plain Java objects and thus first-class citizens, patterns
may also be configured algorithmically by meta-programming in the host system.

In the following presentation, all code samples are in Java\,8, which we assume
the reader is basically familiar with.  We shall take the liberty to add a
keyword \lstinline|partial| for partial type definition {\NEEEEW fragments that
  add up throughout a collection of sources}, borrowed from C\#, in order to focus on
distinct aspects of the APIs according to the flow of discussion.

The \textsf{Paisley} design aims at representing the imperative object-oriented
view on Java data objects faithfully.  Thus it is concerned with the full
spectrum of \emph{operational} semantics of data query operations, of which the
implementation of algebraic semantics is merely a particularly well-behaved
special case.  The basic API is deceptively simple:
\begin{lstlisting}
partial abstract class Pattern<A> {
  public boolean match(A target);
  public boolean matchAgain(); ^\smallskip^
}
\end{lstlisting}
A pattern is an object that can be attempted to \lstinline|match| against some
value \lstinline|target| of the parameter type \lstinline|A|, and will indicate
success by its \lstinline|boolean| return value.  All additional information,
such as extracted pieces of data, needs to be communicated via side effects.
Patterns are potentially non-deterministic; additional matches beyond the
successful first, each with their own observable side effects, can be obtained
by iterating \lstinline|matchAgain| until it fails.  Note that patterns are
required to store the information needed for backtracking as private mutable
state, thus they are reusable sequentially but not concurrently.

The event of a successful match, together with the collection of all
observable side effects, is called a \emph{solution}.
The sequence of all solutions is the primary behavioral semantics of a
pattern.

The single most important pattern class is the \lstinline|Variable|, which can
be bound to data obtained from the target:
\begin{lstlisting}
partial class Variable<A> extends Pattern<A> { ^\smallskip^
  A value; ^\smallskip^
  public boolean match(A target) { value = target;  return true;  }
  public boolean matchAgain()    {                    return false; }
}
\end{lstlisting}
A variable pattern simply matches any target deterministically, and records it
as a side effect.\footnote{\NEEEEW This is the only solution-relevant side
  effect discussed in this paper, but others could be implemented by
  user-defined combinators.} Their power comes from the ability to be nested
inside complex patterns, and hence record selected parts of the overall target
data, under controlled conditions.  Note that variable binding is by ordinary
imperative assignment; there are no declarative concepts such as \emph{single
  assignment} (which would prevent {\NEEEEW transparent} sequential reuse) or
\emph{unification} (which is ill-defined for arbitrary non-algebraic data APIs).

The basic usage template consists of four steps: (1)~allocate pattern variables
to hold results; (2)~construct a complex pattern over the variables; (3)~attempt
one or more matches; (4)~on success, proceed using the result values; see
Fig.~\ref{fig:usage}.

\begin{figure}[t]
  \centering
  \begin{minipage}{.8\textwidth}
    \hrule height \heavyrulewidth
    \lstset{moredelim=**[is][\color{black!50}]{!}{!},moredelim=**[is][\NEEEW]{@}{@},xleftmargin=1em}
    \begin{lstlisting}[gobble=2]
  Variable<V1> v1 = new Variable<>();                       // (1)
  // ...
  Variable<Vn> vn = new Variable<>();                       // (1) ^\smallskip^
  Pattern<A> p = createPattern(v1, ..., vn);                // (2) ^\smallskip^
  if (p.match(target)) !do!                                   // (3a) ^\smallskip^
      doSomething(v1.value, ..., vn.value);                 // (4) ^\smallskip^
    !while (@wantingMore() &&@ p.matchAgain());!               // (3b)
    \end{lstlisting}
    \vspace{-\baselineskip}
    \hrule height \heavyrulewidth
  \end{minipage}
  \caption{Basic usage template for \textsf{Paisley} patterns}
  \label{fig:usage}
\end{figure}

Here \lstinline|createPattern| is problem-specific producer code that may build
on operations from the \textsf{Paisley} library, \lstinline|doSomething| is
arbitrary consumer code that does not need to know about patterns, and the
greyed-out part is optional for the case of exhaustive search of matches for
non-deterministic patterns.  Note that the API is statically type-safe for both
targets and results, and backtracking is subject to explicit imperative control
flow{\NEEEW, including the user-defined condition \lstinline|wantingMore|}.

\subsection{Summary of Features}

This section gives a brief overview of the features of the \textsf{Paisley} core
library.  It is not intended as a detailed or complete introduction, but rather
to convey an intuition about the operational principles and recurring idioms, as
well as the scope of the task of developing a compiler for the \textsf{Paisley}
language.

\newml{\textsf{Paisley} is a \emph{combinatorial language} in the sense of
Schönfinkel and Curry.}
\newml{E}ach primitive
is either a full-fledged pattern that can be used
on its own, or an operator that \newml{builds new patterns from one or more
existing ones.}
The core library can be extended as needed by giving new
implementations (subclasses) of the existing APIs.

\paragraph*{Logic}

The most basic \textsf{Paisley} combinators are \lstinline|both| and
\lstinline|either|, which implement the logical conjunction and disjunction of
patterns, respectively.

The pattern \lstinline|both(p, q)| produces all solutions of \lstinline|q| for
each successive solution of \lstinline|p| in order, both applied to the same
target.  Since \lstinline|q| may observe the variable bindings established by
the successful match for \lstinline|p|, the solution semantics of the combinator
is a \emph{dependent sum} rather than {\NEEEW just} a Cartesian product of the
individual semantics.

The pattern \lstinline|either(p, q)| produces all solutions of \lstinline|p|
followed by all solutions of \lstinline|q|, both applied to the same target.
This is the most straightforward way to introduce non-determinism.  Since
\lstinline|q| is only invoked after solutions for \lstinline|p| are exhausted,
the latter can not observe the former, and the solutions semantics of the
combinator is just the concatenation of the individual semantics.  Note that a
variable can only be considered bound in \newml{each} solution of
\lstinline|either(p, q)| if it is bound by \emph{both} \lstinline|p| and
\lstinline|q|.

\paragraph*{Projections}

Any data access operation that can be reified as an instance \lstinline|f| of
the Java standard interface \lstinline|Function<A, B>|, such as a getter for a
field of type \lstinline|B| from objects of class \lstinline|A|, contravariantly
induces a transform from \lstinline|Pattern<B> p| to
\lstinline|Pattern<A> transform(f, p)| ---
namely, \lstinline|transform(f, p).match(a)| should behave equivalently to
\lstinline|p.match(f.apply(a))|.  This allows patterns operating on parts of a
data structure to be lifted to patterns operating on the whole, by transforming
them with the appropriate access operation.

\paragraph*{Tests}

Any data access operation that can be reified as an instance \lstinline|t| of
the Java standard interface \lstinline|Predicate<A>|, such as a
\lstinline|boolean|-valued getter or an \lstinline|instanceof| test,
induces \lstinline|Pattern<A> guard(t)| --- namely,
\lstinline|guard(t).match(a)| should behave equivalently to
\lstinline|t.test(a)|.  Thus, the pattern matches a target deterministically and
without extra side effects, if and only if the underlying predicate is
satisfied.

\paragraph*{Encapsulated Search}

An important usage of non-deterministic computations embedded in a conventional
deterministic program is \emph{encapsulated search}: locally enumerating all
solutions of a non-deterministic subproblem, without leaking backtracking
control flow to the consumer.  \textsf{Paisley} provides convenience operations
for encapsulating the ubiquitous special case of patterns with a single
variable.  An expression of the form \lstinline|v.bindings(p, a)| enumerates the
values of variable \lstinline|v| for all solutions of \lstinline|p.match(a)|.
Both eager and lazy evaluation are supported:
\begin{lstlisting}
partial class Variable<A> {
  public <B> List<A>      eagerBindings(Pattern<B> pattern, B target);
  public <B> Iterable<A>  lazyBindings (Pattern<B> pattern, B target);
}
\end{lstlisting}

\paragraph*{Pattern Algebra}

For meta-programming with patterns, it would be desirable to be able to
substitute a \lstinline|Variable<B> v| occurring in a \lstinline|Pattern<A> p|
with another \lstinline|Pattern<B> q|.  Since patterns are specified by an
abstract API and in general have no discernible term structure, this is not
straightforward.  If \lstinline|v| is definitely bound in \lstinline|p| however,
we can have the next best thing: an external data-flow composition
\lstinline|v.bind(p, q)| --- namely \lstinline|v.bind(p, q).match(a)| should
behave equivalently to \lstinline|b -> q.match(b)| iterated disjunctively over the
elements of \lstinline|v.lazyBindings(p, a)|.  Note that lazy evaluation ensures
that computations from \lstinline|p| and \lstinline|q| are interleaved in the
expected order~\cite{whyfp}.

Substitution in turn is good enough to define a \lstinline|lambda| operator for
pattern function abstraction.  Considering functions on patterns (\emph{motifs})
as first-class citizens raises the level of abstraction considerably:
\begin{lstlisting}
partial interface Motif<A, B> extends Function<Pattern<A>, Pattern<B>> { } ^\smallskip^
partial class Variable<A> {
  public <B> Motif<A, B> lambda(Pattern<B> body);
}
\end{lstlisting}

Besides the basic composition operations for point-free construction
\newml{(e.g.\ lifted \lstinline!transform()! and \lstinline!guard()!)}, motifs
also provide Kleene \lstinline|star()| and \lstinline|plus()| operators for
full-fledged relational programming~\cite{paisley-xpath}.  {\NEEW These
  operations implement unbounded iteration of a pattern {\NEEEW transparently}
  by lazy cloning, and thus increase the expressive power of \textsf{Paisley}
  considerably.}  See Fig.~\ref{fig:countdown} for a concise
example.

\begin{figure}[tp]
  \centering
  \begin{minipage}{0.85\linewidth}
    \hrule height \heavyrulewidth
    \begin{lstlisting}[gobble=4,mathescape=true,xleftmargin=1em]
    Motif<Integer, Integer> positive    = Motif.guard(n -> n _>_ 0),
                            pred        = Motif.transform(n -> n - 1),
                            countdown  = Motif.star(positive.andThen(pred)); ^\smallskip^
    System.out.println(countdown.eagerBindings(10));
    \end{lstlisting}
    \hrule height \lightrulewidth\medskip

    \quad $\leadsto$ \texttt{[10, 9, 8, 7, 6, 5, 4, 3, 2, 1, 0]} \medskip

    \hrule height \heavyrulewidth
  \end{minipage}
  \caption{Relational programming on numbers}
  \label{fig:countdown}
\end{figure}

\paragraph*{Standard Data Bindings}

The API design of \textsf{Paisley} is modular and open, such that pattern
primitives that bind to actual data APIs can be added as needed.  For
convenience, the core library comes with predefined bindings for some of the
most common Java datatypes: objects (equality, type checks); numbers
(comparison, arithmetic); strings (substrings, regular expressions);
collections and arrays (shape checks, element iteration); XML (DOM trees, XPath
relations).

\subsection{Bottom-Up Optimization}

A major downside of highly generic and reusable combinators is that, without a
specialization framework, their implementation is quite hard to optimize.  By
the very compositional nature of the combinators, the code that implements the
operational semantics of each is a small fragment, and has {\NEEEW hardly any}
metadata about its context that could be exploited for optimization.

We shall take a short detour to demonstrate the optimization potential given by
even the most rudimentary bottom-up context information.  The remainder of this
paper is then the description of a complementary, technologically more
sophisticated solution that also takes the more powerful top-down metadata flow
into account.

The \textsf{Paisley} API specifies a single item of heuristic metadata, namely a
flag that indicates whether a pattern is statically guaranteed to be
deterministic, i.e., not to match any single target more than once:
\begin{lstlisting}
partial class Pattern<A> {
  public boolean isDeterministic();
}
\end{lstlisting}
This information is exploited by the pattern combinator \lstinline|both(p, q)|
that implements the conjunctive sequential combination of patterns \lstinline|p|
and \lstinline|q| (analogous to the Prolog comma operator \texttt{(p, q)}).  If
\lstinline|p| is \emph{not} certainly deterministic, then storage for
backtracking (analogous to a frame of the Prolog choice stack) must be
allocated, for restarting \lstinline|q| for each solution of \lstinline|p|.
Otherwise, both the choice-point storage and the corresponding fragment of a
global backtracking algorithm can be elided.  Note that possible non-determinism
of \lstinline|q| is irrelevant, as it must be realized further down.

The choice between the generic, backtracking implementation and the
optimized, semi-deterministic one is made at pattern construction
time, depending on the value returned by
\lstinline|p.isDeterministic()|.  Figure~\ref{fig:both} depicts both
implementations in horizontal synopsis.  The subpatterns
\lstinline|p|/\lstinline|q| are stored as
\lstinline|left|/\lstinline|right|, respectively.  It is easy to see
that the optimized version is significantly superior in terms of space
and time efficiency, and that this optimization is crucially necessary
for ensuring that \textsf{Paisley} non-determinism does not impose
prohibitive costs where it is not needed.

\begin{figure}[tp]
  \centering
  \hrule height \heavyrulewidth
  \begin{minipage}[t]{0.57\linewidth}
    \begin{lstlisting}[gobble=4,xleftmargin=0.5em]
    private A target_save;
    private boolean left_matched; ^\smallskip^
    public boolean match(A target) {
      return (left_matched = left.match(target))
          && matchNext(target_save = target, false);
    } ^\smallskip^
    public boolean matchAgain() {
      return left_matched
          && matchNext(target_save, true);
    } ^\smallskip^
    private boolean matchNext(A target,
                                 boolean again) {
      if (again ? right.matchAgain()
                : right.match(target))
        return true;
      else
        while (left_matched = left.matchAgain())
          if (right.match(target))
            return true;
      return false;
    }
    \end{lstlisting}
  \end{minipage}%
  \hfill\vrule width \lightrulewidth\hfill
  \begin{minipage}[t]{0.40\linewidth}
    \begin{lstlisting}[gobble=4,xleftmargin=0pt]
    // ^no mutable fields, but^
    // ^assert left.isDeterministic();^ ^\smallskip^
    public boolean match(A target) {
      return left.match(target)
          && right.match(target);
    } ^\smallskip^
    public boolean matchAgain() {
      return
             right.matchAgain();
    }
    \end{lstlisting}
  \end{minipage}
  \hrule height \heavyrulewidth
  \caption{Pattern conjunction, non-deterministic ({\em left}) and
    semi-deterministic ({\em right})}
  \label{fig:both}
\end{figure}

\section{Compiling Paisley}
\label{compile}

The basic mode of \textsf{Paisley} pattern execution is by a modular
interpreter; each object in the graph making up a complex pattern
encapsulates the code and the state variables required for a
particular step of the overall pattern-matching algorithm.  While
elegant and lightweight, this technique has evident limitations regarding
performance.

Fortunately however, combinators have the ideal structure for a
well-known compilation technique, namely \emph{partial evaluation}.
The inputs to each fragment of implementation are clearly
distinguished into two categories of binding time: Combinator
arguments make up the pattern structure, and are bound at pattern
\emph{construction} time; targets are bound at pattern
\emph{application} time.  Thus a pattern may be specialized after
construction, exploiting the information of the former, to obtain the
code of a residual program that just inputs the latter --- that is, an
equivalent monolithic pattern.

{\NEW
Compiling an interpreted language by explicitly controlled partial evaluation of
the interpreter is a ubiquitous and well-proven technique, ultimately haling
back to Futamura's first projection \cite{futamura}, but more recently known as
\emph{staging}\,\cite{metaml}.
}

\subsection{Design of the Paisley Compiler}

The user perspective on \textsf{Paisley} pattern compilation is an extremely
simple API that subsumes interpreted and compiled patterns transparently, and
requires no configuration or global context:
\begin{lstlisting}
partial class Pattern<A> {
  public Pattern<A> compile();
}
\end{lstlisting}
Here \lstinline|p.compile().match(a)| should behave equivalently to
\lstinline|p.match(a)|, although hopefully with less computational overhead,
{\NEEEW as returns on the resources invested in compilation}.
Semantic equivalence implies that \lstinline|p.compile()| shares pattern
variables with \lstinline|p|, but higher-level combinators may have been fused
to a single object, whose code can be executed without internal dynamic function
calls {\NEW and field indirections}, and thus optimized far more aggressively by
the jit compiler.

\begingroup\NEEW

\subsection{Implementation of the Paisley Compiler}


The Java language and virtual machine (JVM) have no native support for partial
evaluation, and are in general not a suitable candidate either, due to their
complex imperative semantics.  Thus \emph{homoiconic} staged meta-programming,
where object and meta code share the same syntax, is not an option.  The JVM
does, however, support dynamic extensions of the code base through class loaders.
Given an expressive JVM bytecode synthesis tool, partial evaluation can be
implemented for well-behaved reified languages, in particular declarative
lightweight EDSLs such as \textsf{Paisley}, with reasonable effort.

We have implemented such a tool based on our \textsf{LLJava}\,\cite{lljava}
framework.  \textsf{LLJava} defines both a low-level JVM programming language
and an abstract bytecode model, and translation tools that can be used as
compiler, disassembler and bytecode manipulation library.  Our experimental new
tool, \textsf{LLJava-live} provides a convenient front-end to the
\textsf{LLJava} bytecode model, particularly tailored to the purpose of modular
synthesis of code for immediate use.  \textsf{Paisley} is its first completed
application.

Generator modules interact with \textsf{LLJava-live} through a
\lstinline|CompilationContext| API that serves both as a source of context (such
as variable bindings) and as a sink for code (such as instructions and scoping
blocks).  Generated code fragments are organized at the intra-method level by
default, and connected in a data-flow network: The enclosing scope of each
fragment denotes $m$ input and $n$ output variables, which are statically typed
and can be realized in bytecode transparently as fields, parameters, temporary
local variables, or arbitrary access code.  For fragments corresponding to
methods, $m$ equals the number of parameters and $n$ equals $1$ or $0$ for a
return value or \lstinline|void|, respectively.

For local data flow, the fragment may read the inputs and must write the outputs
and terminate.  In the process, local variables may be allocated, and nested
fragments inserted and connected.
For non-local data flow, fragments may allocate and share state variables which are
realized as \lstinline|private| fields of the enclosing class.

The virtual instruction set understood by the context comprises both
operand-stack style (\emph{load}/\emph{store}) and register style (\emph{move}).
Basic block generators are passed as \lstinline|Runnable| callbacks, such that
the context can rearrange them as needed.  The code base of the host program can
be referred directly via the standard reification as \lstinline|Class| and
\lstinline|Method| objects.  See Fig.~\ref{fig:live} for an example where a
(highly contrived) code fragment \lstinline|foo| is compiled, including a
subfragment \lstinline|bar|.

\begin{figure}[tp]
  \NEEW
  \centering
  \begin{minipage}{\linewidth}
    \hrule height \heavyrulewidth
    \begin{minipage}{.3\linewidth}
      \begin{lstlisting}
  boolean foo(int n) {
    bar(n + 1);
    return true;
  }

  void bar(int m);
      \end{lstlisting}
    \end{minipage}\vrule width \lightrulewidth%
    \begin{minipage}{.65\linewidth}
      \begin{lstlisting}
  void compileFoo(CompilationContext context) {
    Variable n = context.getInput(0),
             tmp = context.createLocalVariable(int.class);
    context.store(tmp, () -> {
      context.load(n); context.load(1); context.add();
    });
    context.block(asList(tmp), asList(),   // I/O variables
                  () -> compileBar(context));
    context.move(true, context.getOutput(0));
  }
      \end{lstlisting}
      \vspace{-\baselineskip}
    \end{minipage}
    \hrule height \heavyrulewidth
  \end{minipage}
  \caption{Code fragments (\emph{left}) and \textsf{LLJava-live} generator (\emph{right}).}
  \label{fig:live}
\end{figure}

The overall organization of generated code into methods and the API of the
generated class is handled by an application-specific compiler entry point.
\textsf{LLJava-live} provides a generic service for generating the actual
bytecode, loading the class and instantiating it via reflection.

\endgroup

\begingroup\NEW

\paragraph*{Compilation API}

In order to preserve the modularity of \textsf{Paisley}, the compiler is
distributed over the classes that implement pattern combinators, completely
analogous to the interpreter.  Thus, for every method related to interpretation,
we have added a companion method that generates the equivalent code:
\begin{lstlisting}
partial class Pattern<A> {
  protected void compileMatch       (CompilationContext context);
  protected void compileMatchAgain(CompilationContext context);
}
\end{lstlisting}

\begingroup\NEEW
\newml{Calling t}he entry point \lstinline|Pattern.compile()|
\newml{generates} a new subclass of
\lstinline|Pattern| and populates its API methods by invoking each of the
companion methods of the pattern to be compiled with a corresponding context.
In the following, we discuss a few selected issues to be addressed for the
effective compilation of EDSLs in general, and of \textsf{Paisley} in
particular.
\endgroup

\paragraph*{Variable Capture}

As usual in partial evaluation, the program fragments produced by the
construction stage may capture \newml{host language} variables of their context.  For
primitive types, a constant corresponding to the environment value can
simply be injected into the target class.  But capturing references to
live Java objects is another matter.  We use a staged version of the
same technique also employed by the Java compiler for variable
captures in local classes: The target class is
\emph{closure}-converted, that is, captured variables are represented
as \lstinline|private final| fields, and properly initialized with the
environment values when the class is instantiated for proceeding to
the application stage.

\paragraph*{Fallback Strategy: Staged Eta Expansion}

For incremental upgrading of the \textsf{Paisley} core library to compilation,
but also for users who wish to extend the language but not be bothered with
{\NEEEW \textsf{LLJava-live}} code generation, there is a fallback mechanism
that allows any combinator without a specific code generator {\NEEEEW(and its arguments)} to
be embedded in a tree that is compiled as a whole.  This fallback is defined as
the default implementation of code generation methods, which can {\NEEEW either}
be overridden specifically or simply inherited.

The technique is essentially a staged variant of eta expansion, or \emph{reverse
  stubs} in virtual machine terminology: by default, any API method of a pattern
compiles into a call of itself, thus reverting from compiled to interpreted
mode.  This entails the capture of a reference to the original pattern.  As a
special case, pattern variables are always compiled in this way, since their
identity is crucial to the external work flow (see Fig.~\ref{fig:usage}), and
must not be ``optimized'' away {\NEEEEW such that remote interactions via observable side effects are severed.}

\endgroup

\paragraph*{Avoiding Code Explosion}

Partial evaluation frameworks typically draw their power from two related
top-down heuristics: The first is \emph{inlining}, where a function call is
replaced by the function body, specialized by substituting the actual parameter
values for the formal ones.  The second is ``the Trick''\,\cite{trick}, where a
fragment of code depending on an unbound variable with few distinct possible
values, is replaced by a case distinction over the variable, with the original
fragment specialized repeatedly by substituting one possible value per
branch.

Both involve the duplication of code in environments with more bound variables
than the original place of definition, trading the potential for subsequent
simplification for the danger of combinatorial code explosion.  For example in
Fig.~\ref{fig:both}~(\emph{left}), consider the double occurrence of the
inlinable call \lstinline|right.match(target)|, and the parameter variable
\lstinline|boolean again|
that is subject both to inlining globally and to the Trick locally.

Because of the highly self-similar nature of combinator trees, any local
duplication of code can easily lead to exponential growth.  In the context of
the JVM, where the bytecode size of a method is tightly limited to 64\,kiB, and
the resource-constrained verifier and jit compiler are liable to choke on far
less, this becomes a problem very quickly.  Thus duplication of bytecode must be
strictly controlled for the compilation of nestable combinators.

The \textsf{Paisley} compiler has an all-or-nothing policy regarding code
duplication: when the compilation step for any combinator finds that it would
call the same substep more than once, a \lstinline|private| auxiliary method is
created instead, populated once and called from every occurrence.  The decision
whether to inline such methods (where cheap enough) is left to the jit compiler,
which has sophisticated code-size budgeting heuristics anyway.

\subsection{Motif Compilation}

Surprisingly, lifting compilation to the function level, that is from patterns
to motifs, requires hardly any effort.  An obvious naïve solution would be to
compile any motif point-wise:
\begin{lstlisting}
partial interface Motif<A, B> {
  public default Motif<A, B> compile() {
    return p -> this.apply(p).compile();
  }
}
\end{lstlisting}
But this would {\NEEEW redundantly} create a \emph{new} class for every application of a motif.
Fortunately, we can do much better
\newml{by reducing {\NEEEEW the general task} to a clever
treatment of lambda abstractions, \lstinline|v.lambda(p)|, {\NEEEEW that escapes the modular code generation scheme in a substantial but transparent way.}}

Assuming that \lstinline|v| actually occurs in \lstinline|p|, the compilation of
\lstinline|p| will
{\NEEEEW include} 
the staged eta expansion of \lstinline|v|.
Hence \lstinline|v| will occur in the environment of the compiled closure.  All
we need to do is to defer the actual constructor call for the closure, and
return a motif that calls the constructor when applied, substituting its
argument for \lstinline|p| in the environment.  In short,
\lstinline|v.lambda(p).compile().apply(q)| should behave equivalently to
\lstinline|p.compile()|, except that the latter's environment reference to
\lstinline|v| is rerouted to \lstinline|q|.

No other motif combinator needs to be implemented manually.  Any
complex motif \newml{\lstinline!m!} can be compiled monolithically by
instead compiling its eta expansion,
\newml{\lstinline!m.etaExpand().compile()!}{\NEEEEW, where the above
  procedure can be applied to the body}.

\begin{lstlisting}
partial interface Motif<A, B> {
  public default Motif<A, B> etaExpand() {
    Variable<A> x = new Variable<>();
    return x.lambda(this.apply(x));
  }
}
\end{lstlisting}
The only catch is that the variable \lstinline|x| is {\NEEEEW naturally}
considered deterministic in
the construction-time analysis of \lstinline|p|, as discussed above.  Thus for
non-deterministic patterns \lstinline|q| backtracking glue code needs to be
inserted. 
{\NEEEEW The implementation of \lstinline|compile()| for eta-expanded motifs deals with this transparently.}

\section{Evaluation}
\label{eval}

We evaluate the performance of the \textsf{Paisley} compiler and its results by
reiterating previously published benchmarks of (interpreted) \textsf{Paisley}
applications.\footnote{All results reported here have been obtained on the
  same test equipment, namely a Core~i7-5600U\,@\,2.60\,GHz CPU with 16\,GiB of
  RAM, running CentOS~Linux~7 and OpenJDK~8u202.}

\subsection{Cryptarithmetic Puzzles}

In \cite{crypt} we demonstrated the use of \textsf{Paisley} for embedded
logic programming by considering \emph{cryptarithmetic puzzles}.  Given a
natural number $b$, an injective mapping of letters to values in $\{0, \dots,
b-1\}$ induces a $b$-adic notation of natural numbers disguised as words.  A
puzzle is a sum equation of $n$ words, and the solutions are the mappings that
satisfy the equation.  The classic example is $SEND + MORE = MONEY$, with $b =
10$ and $n = 2$, which has the unique solution $O = 0$, $M = 1$, $Y = 2$, $E =
5$, $N = 6$, $D = 7$, $R = 8$, and $S = 9$ \cite{send}.

Our approach to solving cryptarithmetic puzzles with \textsf{Paisley} is based
on one pattern variable for each letter, and the set of possible digits as the
target object.  Various generic non-deterministic combinators from the
\textsf{Paisley} library span the search tree, and a few problem-specific
\emph{constraint} patterns prune it.  (Constraint patterns do not examine the
target object, but the bindings of variables, exploiting the dependent nature of
the \lstinline|both| combinator.)

In \cite{crypt} we considered three increasingly sophisticated search-plan
construction algorithms for arbitrary cryptarithmetic puzzles:
\begin{enumerate}
\item A \emph{naïve} generate-and-test strategy that exhausts the Cartesian
  space of variable bindings by brute force, and checks the injectivity and
  arithmetic constraints for each at the very end.
\item A strategy that exploits \emph{injectivity} by inserting pair-wise
  inequality constraints for bound variables as early as possible.
\item A strategy that additionally exploits \emph{modular arithmetic} by binding
  variables in right-to-left order of occurrence, inserting approximative checks
  for the sum modulo $b^k$, for increasing $k$, as early as possible.
\end{enumerate}

\begin{table}[tp]
  \caption{Solving the SEND+MORE=MONEY puzzle with \textsf{Paisley} patterns.}
  \label{tab:crypt}
  \centering
  \def\arraystretch{1.2}\small\tabcolsep=0.5em
  \begin{tabular}{l@{\quad}rrc@{\quad}rrrr}
    \toprule
    \multirow{2}{*}{\textbf{Strategy}} & \multicolumn{2}{c}{\textbf{Run Time}} & \multirow{2}{*}{\textbf{Speedup}} & \multicolumn{4}{c}{\textbf{Compilation}}
    \\
    & \textbf{interp.} & \textbf{compiled} && \textbf{time} & \textbf{bytes} & \textbf{flds} & \textbf{mths}
    \\ \midrule
    \textbf{naïve} & 4\,029\,ms & 3\,530\,ms & 1.14 & 17.8\,ms & 8\,339 & 35 & 29
    \\
    \textbf{injective} & 636\,ms & 279\,ms & 2.28 & 23.4\,ms & 21\,932 & 91 & 85
    \\
    \textbf{modular} & 1\,719\,\textmu s & 813\,\textmu s & 2.11 & 23.5\,ms & 23\,892 & 99 & 93
    \\ \bottomrule
  \end{tabular}
\end{table}

We have re-run the cryptarithmetic puzzle solver application, using
out-of-the-box compilation support for all generic combinators of the
\textsf{Paisley} core library, but strictly no additional problem-specific
generator code.  Table~\ref{tab:crypt} summarizes our benchmarking results.
For each strategy the following data are given:
\begin{itemize}
\item run times of the original pattern and its compiled variant,
  and their ratio;
\item times for compilation, including bytecode generation, class loading and
  verification and object initialization;
\item size of generated class, measured in overall bytes, number of state fields
  and matching-related methods (\lstinline|match|, \lstinline|matchAgain| and
  their auxiliaries).
\end{itemize}

All reported times are wall-clock times, each obtained with
\lstinline|System.nanoTime()| precision, as the median of a specific, suitably
large number of iterations to allow for jit compiler warm-up.  {\NEEW See
  section~\ref{conclusion} for further discussion.}

\subsection{Document Object Model Navigation with XPath}

\begingroup\NEW

XPath\,\cite{xpath10} is a declarative non-deterministic domain-specific
language for navigation in XML document trees, suitable for embedding in various
more high-level XML technologies such as XQuery
\ and XSLT
.  In \cite{paisley-xpath}, we demonstrated how a straightforward translation of
XPath\,1.0 abstract syntax to \textsf{Paisley} motifs yields a lightweight lazy
XPath execution engine, which is not only highly educational, but even in
interpreted form competes well against the heavyweight XML tools shipped with
the Java platform.  As benchmarks, we used a selection of test cases from the
XMark\,\cite{xmark} suite\newml{, see Table~\ref{tab:xmark}.}

\begin{table}
   \newml{\caption{Executing XPath queries from the XMark suite with \textsf{Paisley} patterns.}}
  \label{tab:xmark}
  \label{tab:xpath}
  \centering
  \def\arraystretch{1.2}\small\tabcolsep=0.5em
  \begin{tabular}{lrrrcccr}
    \toprule
    \textbf{Test} & \multicolumn{7}{l}{\textbf{XPath Expression}}
    \\ \midrule
    \textbf{Q00} & \multicolumn{7}{l}{\texttt{//node()}}
    \\
    \textbf{Q01} & \multicolumn{7}{l}{\texttt{/site/open\_auctions/open\_auction/bidder[1]/increase/text()}}
    \\
    \textbf{Q06} & \multicolumn{7}{l}{\texttt{//site/regions//item}}
    \\
    \textbf{Q15} & \multicolumn{7}{l}{\def\arraystretch{1}\tabcolsep=0pt\begin{tabular}[t]{l}\texttt{/site/closed\_auctions/closed\_auction/annotation/description/}\\\quad\texttt{parlist/listitem/parlist/listitem/text/emph/keyword/text()}\end{tabular}}
    \\
    \textbf{Q16} & \multicolumn{7}{l}{\def\arraystretch{1}\tabcolsep=0pt\begin{tabular}[t]{l}\texttt{/site/closed\_auctions/closed\_auction[annotation/description/}\\\quad\texttt{parlist/listitem/parlist/listitem/text/emph/keyword/text()]}\end{tabular}}
    \\ \bottomrule
%
    \toprule
    \multirow{2}{*}{\textbf{Test}} & \multicolumn{3}{c}{\textbf{Run Time}} & \multicolumn{3}{c}{\textbf{Overhead}} & \multirow{2}{*}{\textbf{Solutions}}
    \\
    & \textbf{interp.} & \textbf{comp.} & \textbf{baseline} & \textbf{interp.} & \textbf{comp.} & \textbf{speedup}
    \\ \midrule
    \textbf{Q00} & 99.56\,ms & 64.05\,ms & 13.71\,ms & 6.26 & 3.67 & 1.71 & 1\,877\,979
    \\
    \textbf{Q01} & 11.16\,ms & 6.78\,ms & 5.44\,ms & 1.05 & 0.25 & 4.27 & 4\,310
    \\
    \textbf{Q06} & 162.85\,ms & 84.52\,ms & 62.11\,ms & 1.62 & 0.36 & 4.50 & 8\,700
    \\
    \textbf{Q15} & 7.13\,ms & 4.07\,ms & 3.62\,ms & 0.97 & 0.12 & 7.80 & 68
    \\
    \textbf{Q16} & 9.43\,ms & 4.55\,ms & 3.92\,ms & 1.41 & 0.16 & 8.75 & 59
    \\ \bottomrule
  \end{tabular}
\end{table}

We have re-run the tests, using compilation support for all generic combinators
of the \textsf{Paisley} core library, as well as for bindings to the standard
Java XML DOM.  Table~\ref{tab:xpath} summarizes our benchmarking results.  For
each test the following data are given:

\begin{itemize}
\item run times of the original motif and its compiled variant;
\item the baseline run time of a hand-coded eager traversal algorithm that
  efficiently implements that particular XPath expression;
\item the relative overhead of the interpreted and compiled \textsf{Paisley}
  variants over the baseline, and their ratio;
\item the number of solutions
\end{itemize}
\dots\ in a fixed pseudo-random input document, generated by a tool supplied by
the authors of XMark.\footnote{The official home page is no longer online, but
  retrievable from
  \url{https://web.archive.org/web/20070810005114/http://www.xml-benchmark.org/}.}
All reported times are obtained as above.  The results show that the generic
\textsf{Paisley} implementation of XPath expressions approximates the
performance of specific one-off Java implementations gracefully.

The overhead is noticeable in case Q00, where a trivial query basically matches
all nodes, and thus yields a huge number of solutions.  Here the cost of lazy
backtracking, as opposed to eager traversal, has an impact that can not be
compensated {\NEEEW fully} by our compilation technique.  On the upside, the
lazy search can be suspended arbitrarily after each solution, at no additional
cost.
For the other cases, where significant amounts of traversal take place between
solutions, the \textsf{Paisley} overhead is moderate.  Furthermore it can be
improved to near insignificance by compilation, such that the costs of actually
calling into the target data API completely dominate.

\endgroup

\section{Conclusion}
\label{conclusion}

\begingroup\NEEW

We have demonstrated how staged compilation can improve the performance of
\textsf{Paisley}, a modularly interpreted combinator EDSL par excellence.  The
compiler mirrors the structure of the interpreter and generates bytecode that
can be immediately loaded and eventually jit-compiled by the JVM.  Compiled
and interpreted \textsf{Paisley} interface transparently in both directions, and
dealing with compilation is completely optional for user extensions.
The approach is generally suitable also for accelerating any other declarative EDSL.

Benchmarks indicate that the speedup by compilation is significant, even for
legacy applications, and can approximate hand-written data query code.  We
foresee that long-running applications with complex internal data models, such
as information systems and document servers, could benefit the most from this
technology.  This is because their usage mode fits the assumptions of staged
compilation perfectly: \emph{construct early, reuse often}.

In a multi-stage pipeline such as the jit-compiled JVM, there is more to
consider than just the run time of the compilation step.  For pattern compilation
to pay off in the end, the compiled patterns must be (re-)used often enough for
the jit compiler to consider them worthwhile for machine code generation.
Otherwise they are executed compiled at the level of the embedded language
\textsf{Paisley}, but interpreted at the level of the host, in contrast to the
original patterns for which the situation is the converse.  Thus one-off
applications such as the cryptarithmetic puzzles are purely academic, and for
more heterogeneous realistic applications empirical validation is required.

\endgroup

\subsection{Related Work}


{\NEEEEW Many different approaches to pattern matching in Java exist.
  We have already compared our approach to the most significant ones,
  in particular the historically relevant
  \textsf{JMatch}\,\cite{jmatch-padl} in previous papers
  \cite{paisley-atps,paisley-kawa}.  More modern, quasi-algebraic
  solutions, such as \textsf{adt4j}\footnote{\url{https://github.com/sviperll/adt4j}} or
  \textsf{derive4j}\footnote{\url{https://github.com/derive4j/derive4j}},
  do not properly address object-oriented data abstraction and non-determinism, the
  focus of \textsf{Paisley} in general, or compilation, the focus of
  the present paper in particular.  }

On the JVM, the Scala language supports non-algebraic pattern matching via
dedicated syntax and the magic method \lstinline|unapply|.  As a core part of the
language and its compiler, this mechanism is much more tightly integrated than
\textsf{Paisley} can ever hope to be, and naturally compiles both predefined and
custom pattern code.  But the comparison is not exactly fair, as Scala patterns
are neither non-deterministic, nor point-free, nor dynamically
meta-programmable.

\begingroup\NEW
A very recent work \cite{parsing} on parser generation has inspired us to
complete the work presented here.  They also improve the performance of a
combinator language, often drastically, by intermediate compilation of a
construction stage.  Their approach, like ours, combines the benefits of
bottom-up heuristic metadata (a variant of $\mathrm{LL}(1)$ analysis) with those
of top-down code specialization.  However, the MetaOCaml host language framework
they use is markedly different in nature: On the one hand, it natively supports
staged meta-programming, for which we have had to build a custom tool onto
Java's dynamic bytecode loading.  On the other hand, OCaml does not have the
benefit of a jit compiler that could optimize both combinators and generated
code heuristically, which makes their compilation stage proportionally even more
effective.
\endgroup





\begin{thebibliography}{99}

\small

\bibitem{send} \hypertarget{cite.0@send}{}%
  H.\ E.\ Dudeney. In: \textit{Strand Magazine}~68 (July 1924),
  pp. 97, 214.

\bibitem{whyfp} \hypertarget{cite.0@whyfp}%
  J. Hughes.  ``Why Functional Programming Matters''.  In:
  \textit{Comput.\ J.}~32.2 (1989), pp.~98--107. \textsc{doi:}
  \texttt{\href{https://dx.doi.org/10.1093/comjnl/32.2.98}{10.1093/comjnl/32.2.98}}.

\bibitem{trick} \hypertarget{cite.0@trick}%
  O. Danvy, K. Malmkjær, and J. Palsberg.  ``Eta-Expansion Does The
  Trick''.  In: \textit{ACM Trans.\ Program.\ Lang.\ Syst.}~18.6
  (1996), pp.~730--751. \textsc{doi:}
  \texttt{\href{https://dx.doi.org/10.1145/ 236114.236119}{10.1145/
      236114.236119}}.

\bibitem{xpath10} \hypertarget{cite.0@xpath10}%
  J.\ Clark and S.\ DeRose.  \textit{XML Path Language
    (XPath) Version 1.0}.  W3C.  \textsc{url:}
  \url{https://www.w3.org/TR/1999/REC-xpath-19991116/}, 1999.

\bibitem{futamura} \hypertarget{cite.0@futamura}%
  Y.\ Futamura.  ``Partial Evaluation of Computation Process-- An
  Approach to a Compiler-Compiler. (Revised Reprint)''.  In:
  \textit{Higher-Order and Symbolic Computation}~12 (1999),
  pp.~381--391.  \textsc{doi:}
  \texttt{\href{https://dx/doi.org/10.1023/A:1010095604496}{10.1023/A:1010095604496}}.

\bibitem{metaml} \hypertarget{cite.0@metaml}%
  W.\ Taha and T.\ Sheard.  ``MetaML and multi-stage programming with
  explicit annotations''. In: \textit{Theoretical Computer
    Science}~248.1 (2000), pp.~211--242.  \textsc{doi:}
  \texttt{\href{https://dx.doi.org/10.1016/S0304-3975(00)00053-0}{10.1016/S0304-3975(00)00053-0}}.

\bibitem{xmark} \hypertarget{cite.0@xmark}%
  A.\ Schmidt et al.  ``XMark: A Benchmark for XML Data
  Management''. In: \textit{Proc.\ 28th VLDB}.  Morgan Kaufmann, 2002,
  pp.~974--985. \textsc{url:}
  \url{http://www.vldb.org/conf/2002/S30P01.pdf}.

\bibitem{jmatch-padl} \hypertarget{cite.0@jmatch-padl}%
  J.\ Liu and A.\ C.\ Myers.  ``JMatch: Iterable
  Abstract Pattern Matching for Java''.  In: \textit{Proc.\ 5th PADL}.
  Vol.~2562.  LNCS.  Springer-Verlag, 2003.  \textsc{doi:}
  \texttt{\href{https://dx.doi.org/10.1007/3-540-36388-2_9}{10.1007/3-540-36388-2\_9}}.

\bibitem{mowp} \hypertarget{cite.0@mowp}%
  B.\ Emir, M.\ Odersky, and J.\ Williams.  ``Matching
  Objects with Patterns''.  In: \textit{Proc.\ 21st ECOOP}.  Ed.\ by
  E.\ Ernst.  Vol.~4609.  LNCS.  Springer-Verlag, 2007, pp.~273--298.
  \textsc{doi:}
  \texttt{\href{https://dx.doi.org/10.1007/978-3-540-73589-2_14}{10.1007/978-3-540-73589-2\_14}}.

\bibitem{paisley-icmt} \hypertarget{cite.0@paisley-icmt}%
  B.\ Trancón y Widemann and M.\ Lepper.
  ``Paisley: pattern matching a ` la carte''.  In: \textit{Proc.\ 5th
    ICMT}.  Vol.~7307.  LNCS.  Springer-Verlag, 2012,
  pp.~240--247. \textsc{doi:}
  \texttt{\href{https://dx.doi.org/10.1007/978-3-642-30476-7_16}{10.1007/978-3-642-30476-7\_16}}.

\bibitem{paisley-atps} \hypertarget{cite.0@paisley-atps}%
  B.\ Trancón y Widemann and M.\ Lepper.  ``Paisley: A Pattern
  Matching Library for Arbitrary Object Models''.  In:
  \textit{Software Engineering 2013, Workshopband}.  Vol.~215.  LNI.
  Gesellschaft für Informatik, 2013, pp.~171--186.  \textsc{url:}
  \url{http://www.se2013.rwth-aachen.de/downloads/proceedings/SE2013WS.pdf}.

\bibitem{crypt} \hypertarget{cite.0@crypt}%
  B.\ Trancón y Widemann and M.\ Lepper.  ``Some Experiments on
  Light-Weight Object-Functional-Logic Programming in Java with
  Paisley''.  In: \textit{Declarative Programming and Knowledge
    Management}.  Vol.~8439.  LNCS.  Springer-Verlag, 2014,
  pp.~218--233.  \textsc{doi:}
  \texttt{\href{https://dx.doi.org/10.1007/978-3-319-08909-6_14}{10.1007/978-3-319-08909-6\_14}}.

\bibitem{paisley-xpath} \hypertarget{cite.0@paisley-xpath}%
  B.\ Trancón y Widemann and M.\ Lepper.  ``Interpreting XPath by
  Iterative Pattern Matching with Paisley''.  In: \textit{Proc.\ 23rd
    WFLP}.  Vol.~1335.  CEUR-WS.org, 2015, pp.~108--124. \textsc{url:}
  \url{http://ceur-ws.org/Vol-1335/wflp2014_paper1.pdf}.

\bibitem{paisley-kawa} \hypertarget{cite.0@paisley-kawa}%
  B.\ Trancón y Widemann and M.\ Lepper.  ``A Practical Study of
  Control in Objected-Oriented–Functional–Logic Programming with
  Paisley''.  In: \textit{Proc.\ 24th WFLP}.  Vol.~234.  EPTCS.  2016,
  pp.~150--164.  \textsc{doi:}
  \texttt{\href{https://dx.doi.org/10.4204/EPTCS.234.11}{10.4204/EPTCS.234.11}}.

\bibitem{lljava} \hypertarget{cite.0@lljava}%
  B.\ Trancón y Widemann and M.\ Lepper.  ``LLJava: Minimalist
  Structured Pro- gramming on the Java Virtual Machine''.  In:
  \textit{Proc.\ 13th PPPJ}.  ACM, 2016.  \textsc{doi:}
  \texttt{\href{httsp://dx.doi.org/10.1145/2972206.2972218}{10.1145/2972206.2972218}}.

\bibitem{parsing} \hypertarget{cite.0@parsing}%
  N.\ Krishnaswami and J.\ Yallop.  ``A Typed, Algebraic Approach to
  Parsing''.  In: \textit{Proc.\ 40th PLDI}.  ACM, 2019, pp.~379--393.
  \textsc{doi:}
  \texttt{\href{https://dx.doi.org/10.1145/3314221.3314625}{10.1145/3314221.3314625}}.

\end{thebibliography}
\end{document}